\documentclass{IAU-TrA}
\usepackage{graphicx}      

\title[ATOMIC AND MOLECULAR DATA] 
{}

\author[DIVISION~B / COMMISSION~14 / WORKING GROUP] 
{ATOMIC DATA }

\pubyear{2015}
\volume{Volume}
\pagerange{001--008}
\jname{Transactions IAU, Volume XXIXA}
\editors{Thierry Montmerle, ed.}
\begin{document}

\maketitle
                    
{\bf

\large
\noindent
DIVISION B / COMMISSION~14 / WORKING GROUP                    \\ 
ATOMIC DATA                                                      \\
                                                                                  
\normalsize                       

\begin{tabbing}
\hspace*{45mm}  \=                                 \kill
CHAIR           \> Gillian Nave                    \\
CO-CHAIRS       \> Sultana Nahar, Gang Zhao        \\
\end{tabbing}
        
\vspace{3mm}

\noindent
TRIENNIAL REPORT 2011-2015            
}
                                    

This report summarizes laboratory measurements of atomic wavelengths,
energy levels, hyperfine and isotope structure, energy level
lifetimes, and oscillator strengths. Theoretical calculations of
lifetimes and oscillator strengths are also included. The bibliography
is limited to species of astrophysical interest. Compilations of
atomic data and internet databases are also included. Papers are
listed in the bibliography in alphabetical order, with a reference number in the
text. 

Comprehensive lists of references for atomic spectra can be found in the NIST 
Atomic Spectra Bibliographic Databases {\tt http://physics.nist.gov/asbib}.

\section{Energy levels, wavelengths, line classifications, and line structure}

Major analyses of wavelengths, energy levels and line classifications have been 
published for {\bf Cr~II}[\cite{18847EL,18287EL}], {\bf Fe~II}[\cite{18349EL}], 
and {\bf V~II}[\cite{18516EL}] in the past three years. Laboratory wavelengths 
and line identifications have been published for coronal lines observed in 
spectra from the Solar Dynamics Observatory in the region 193~\AA\ to 218~\AA\ 
and around 131~\AA\ [\cite{18800EL, 18918EL, 18679EL}]. Wavelengths and 
identifications of additional coronal lines between 170~\AA\ and 291~\AA\ have 
been measured using the Hinode EUV Imaging Spectrometer [\cite{18012EL}]. 

Additional publications of wavelengths, energy levels, line classifications, 
hyperfine structure (HFS), and isotope structure (IS) include:
                    
{\bf Al~II} (HFS):  [\cite{9060TP}],
{\bf Ar~IX}:  [\cite{9386TP}],
{\bf B~II}:  [\cite{18818EL}],
{\bf Br~III}:  [\cite{19007EL}],
{\bf Cl~I}:  [\cite{18583EL}],
{\bf Co~II} (forbidden):  [\cite{9217TP}],
{\bf Dy~I}:  [\cite{18985EL}],
{\bf Eu~I} (HFS,IS): [\cite{18578EL,17830EL,18738EL,18856EL}],
{\bf Fe~VIII-IX}:[\cite{17936EL}],
{\bf Fe~VIII-XVI}: [\cite{18266EL}],
{\bf Fe~XI}: [\cite{18347EL}],
{\bf Fe~XIII}: [\cite{17719EL}],
{\bf Fe~XVI}: [\cite{17922EL}],
{\bf Fe~XVII}: [\cite{18046EL}],
{\bf Fe~XVIII}: [\cite{18390EL}],
{\bf Fe~XVIII-XXV}: [\cite{18627EL}],
{\bf Gd~I} (IS,HFS): [\cite{17824EL}],
{\bf I~I}:  [\cite{19106EL}],
{\bf K~I}:  [\cite{9040TP}],
{\bf K~III}:  [\cite{18870EL}],
{\bf La~I}: [\cite{18876EL}],     
{\bf Li~I}:  [\cite{18475EL}],
{\bf Mg~VII-VIII}:  [\cite{18285EL}],
{\bf Mn~I} (HFS): [\cite{18980EL,18617EL}],
{\bf Mn~II} : [\cite{19113EL}],
{\bf N~III}:  [\cite{18714EL}],
{\bf Na~I}:  [\cite{9184TP}],
{\bf Nb~I}: [\cite{18958EL,18957EL}],
{\bf Nb~I} (HFS): [\cite{17718EL}],
{\bf Nd~II} (HFS): [\cite{18959EL}],
{\bf Ne~IV} :[\cite{17865EL}],
{\bf Ni~I}:  [\cite{18021EL,18193EL}],
{\bf O~I}:  [\cite{18459EL}],
{\bf O~VI}: [\cite{18235EL}],                    
{\bf P~II}:  [\cite{18070EL}],
{\bf Pr~I} (HFS): [\cite{18541EL,17789EL,18249EL}],
{\bf Pr~II}: [\cite{17987EL}],
{\bf Pr~II} (HFS): [\cite{18334EL}],
{\bf Ru~I} (IS,HFS): [\cite{18672EL}],
{\bf S~I}:  [\cite{18070EL}],
{\bf S~VII-XIV}:  [\cite{18322EL}],
{\bf S~VIII}: [\cite{9039TP}],
{\bf Si~IV}:  [\cite{19014EL}],
{\bf Se~III}:  [\cite{17867EL}],
{\bf Sm~I}: [\cite{9218TP}],
{\bf Sn~II}: [\cite{18891EL}],
{\bf Sn~III}: [\cite{9068TP}],
{\bf Ta~II}: [\cite{18919EL}],
{\bf Te~III}: [\cite{17749EL}],
{\bf Ti~X}:  [\cite{9206TP}],
{\bf Ti~IV}:  [\cite{19014EL}],
{\bf Tl~I} (IS,HFS): [\cite{18712EL}],   
{\bf V~I} (HFS):  [\cite{17786EL,19038EL,18877EL}],
{\bf V~II} (forbidden): [\cite{9217TP}],
{\bf V~II} (HFS):  [\cite{19035EL,18027EL}],
{\bf Zn~I}: [\cite{9261TP}]
{\bf Zn~II}:  [\cite{19014EL}].

The references included here for elements heavier than Ni (Z$>$28) are limited
to the first three spectra only, these data being of most interest for
astronomical  spectroscopy.                                                            
                                                                   
Analyses of neutral through doubly-ionized iron-group spectra using Fourier 
transform spectroscopy (FTS) and grating spectroscopy are underway 
at the National Institute of Standards and Technology (NIST), 
USA and Imperial College London, UK. Analysis of moderately ionized species 
(III-VIII) is being done at NIST, USA; Observatoire de Paris-Meudon, France; the 
Institute of Spectroscopy in Troitsk, Russia; Centro Investigaciones Opticas, La 
Plata, Argentina; and Aligarh Muslim University, India. Measurements of HFS and 
IS using both FTS and laser spectroscopy are in progress at Istanbul University, 
Turkey; Lund  University, Sweden; Graz University of Technology, Austria; 
Instituut voor Kern- en Stralingsfysica, Leuven, Belgium; Bhabha Atomic Research 
Centre, India; Pozna\'n University of Technology, Poland; University of 
Birmingham, UK; ISOLDE (international collaboration including Switzerland, 
Germany, France, UK, Russia, Belgium, Spain, Portugal, and Japan); and York 
University, Toronto, Canada. 

Studies of more highly-ionized elements are being done using electron beam ion 
traps (EBIT) at NIST, USA;  Lawrence Livermore National Laboratory, USA; 
Heidelberg, Germany; Shanghai, China; with an accelerator in Beijing, China; and 
at the National Institute for Fusion Science, Japan. 

Theoretical calculations of energy levels, oscillator strengths, HFS, IS, 
photoionization, and collisional data are currently being performed at NIST, 
USA; Ohio State University, USA; Queens University Belfast, Northern Ireland; 
and the University of Lund, Sweden.      

\section{Wavelength standards}

The thorium-argon hollow cathode lamp is an important calibration source for
ground-based astronomical spectrographs and new Ritz wavelengths for
19874 thorium lines measured using FTS have been published [\cite{18702EL}]. In 
the infrared, uranium-neon hollow cathode lamps have advantages over
thorium-argon lamps and an atlas based on FTS for the U/Ne lamp between 850 and 
4000 nm has been published [\cite{17911EL}]. The calibration of these FTS
measurements included reference lines in U and Th measured using optogalvanic 
spectroscopy [\cite{18154EL}]. Additional Ritz wavelengths based on FTS have 
been published for Fe~II, Mg~I-II, Cr~II, Ti~II, Mn~II, Ni~II, and Zn~II
[\cite{17870EL}]. Wavenumbers and pressure shifts in Ar have been 
measured using laser spectroscopy [\cite{18086EL}].
                               
In more highly-charged ions, laser spectroscopy measurements of the 
$\mathrm 2p\ ^2P_{3/2} - ^2P_{1/2}$ forbidden transition in Ar XIV has been 
performed in an EBIT [\cite{17924EL}]. High-accuracy 
wavelength standards in the 3~keV region have been published for lines of 
S XIII-XV, Cl XIV-XV, and Ar XV-XVII [\cite{18507EL}].

\section{Oscillator strengths}

The transition-probability data in the references in section \ref{TP_refs}
were obtained by both theoretical and experimental methods. The
references included here for elements heavier than Ni (Z$>$28) are limited to 
the first three or four spectra only. For Fe II, the set of critically evaluated 
oscillator strengths has been greatly expanded in the NIST Atomic Spectra 
Database ({\tt http://physics.nist.gov/asd}) and now contains data for 6700  
Fe~II lines from 92~nm to 87~$\mu$m. 
Extensive sets of oscillator strengths of fine structure levels
(n$\leq$10) are being calculated with the Breit-Pauli R-matrix (BPRM) method
by S. Nahara, Ohio State Univ, USA (see NORAD database in section \ref{dbase}).
  
\section{Photoionization cross-sections}

Many of the theoretical papers in the references also include data on
photoionization and collisional cross-sections. Additional references for both experimental and calculated cross-sections are as follows:

\begin{minipage}[t]{2in}
 Al V:   \cite{18715EL} \\  
 Al X:   \cite{18708EL} \\  
 Ar VI:   \cite{18668EL} \\ 
 Ar IX: \cite{18497EL} \\   
 B II:  \cite{18818EL} \\   
 Be I:   \cite{18869EL} \\  
 C II:   \cite{18991EL} \\  
 C IV:   \cite{19186EL} \\  
 Ca II:   \cite{18216EL} \\ 
 Ca XI:   \cite{18350EL} \\ 
 Cl I:   \cite{18583EL,19191EL} \\      
 Cu XX:   \cite{18280EL} \\         
\end{minipage}
\begin{minipage}[t]{2in}
 Fe II:  \cite{17927EL} \\       
 Fe X: \cite{18300EL} \\         
 Fe XV:  \cite{19046EL} \\            
 Fe XVII: \cite{18497EL} \\      
 Ga I:   \cite{18498EL} \\       
 Ge I:   \cite{18325EL} \\       
 K III:   \cite{18870EL} \\      
 Kr II:  \cite{18017EL,18575EL} \\          
 N III:  \cite{18714EL} \\  
 N IV-V:   \cite{18552EL} \\
 N IV : \cite{18253EL} \\   
 Ne VII : \cite{18253EL} \\ 
\end{minipage}
\begin{minipage}[t]{2in}
Ne VIII:  \cite{18363EL} \\  
Ni I:   \cite{18021EL} \\    
O II-III:   \cite{19168EL} \\
P III-IV:   \cite{19053EL} \\
S IX:   \cite{19158EL} \\    
S V:   \cite{18560EL} \\     
Se II: \cite{17988EL} \\     
Si IX:   \cite{19158EL} \\   
Sr I:   \cite{18155EL} \\    
Ti XIX:   \cite{18854EL} \\  
Ti XX:   \cite{19056EL} \\   
\end{minipage} 
                                                    
\section{Compilations, Reviews, Conferences}

Major compilations of wavelengths, energy levels or oscillator strengths 
have been published for the following:                
{\bf Ag~II}: [\cite{18362EL}],   
{\bf Cr~I-II}: [\cite{18327EL,18847EL}],  
{\bf Fe~II}:[\cite{18349EL}],   
{\bf Fe~V}: [\cite{18718EL}],
{\bf In~II}: [\cite{18318EL}],                                            
{\bf Mn~II}: [\cite{9140TP}],
{\bf Ne~IV}: [\cite{17865EL}],
{\bf Sn~II}: [\cite{18891EL}],
{\bf Sr~II}: [\cite{8986TP}],
{\bf Ti~I-II}: [\cite{17885EL}].
A summary of the methods used for the critical evaluation of atomic data is 
given in [\cite{18395EL}].                             
                                            
Papers on atomic spectroscopic data are included in the proceedings of
the 11th International Conference on Atomic Spectra and Oscillator
Strengths [\cite{ASOS}] and the 8th International Conference on Atomic and
Molecular Data and their Applications [\cite{ICAMDATA}]. Additional meetings 
including papers on atomic data include the meetings of the Laboratory 
Astrophysics Division (LAD) of the American Astronomical Society (AAS), the 
International Conference on Atomic Processes in Plasmas, the Congress of the 
European Group on Atomic Systems, the International Conference on Phenomena in 
Ionized Gases; and the meeting of the Division of Atomic, Molecular and Optical 
Physics of the American Physical Society.   

\section{Databases}\label{dbase}

The following databases of atomic spectra at NIST 
have received significant updates since the last triennial report:
                                                 
\begin{description}
\item{{\bf NIST Atomic Spectra Database, Version 5 (Sept 2014):} \\}
{\tt http://physics.nist.gov/asd } contains critically compiled data on wavelengths, energy levels and oscillators strengths.
\item{{\bf NIST Atomic Spectra Bibliographic Databases:} (Frequently updated)}\\   
{\tt http://physics.nist.gov/asbib} 
consists of three databases of publications on atomic transition
probabilities, atomic energy levels and spectra, and atomic spectral
line broadening.                   
\end{description}  

Additional on-line databases including significant quantities of
atomic data include: 

\begin{description}
\sloppy
\item{{\bf NIST Database of Basic Atomic Spectroscopic Data, Version 1.1.3,}            
(Nov 2013)}                   
{\tt http://physics.nist.gov/handbook}
Provides a selection of the most important and frequently used atomic 
spectroscopic data in an easily accessible format. 

\item{{\bf The MCHF/MCDHF Collection on the Web, Version 2} (Sept 2010)} 
(C.Froese Fischer \textit{et al}.) at {\tt http://nlte.nist.gov/MCHF/index.html}
contains results of multi{}-configuration Hartree-Fock (MCHF)
or Dirac-Hartree-Fock (MCDHF) calculations for hydrogen and 
Li{}-like through Ar{}-like ions, mainly for  Z $\leq$ 30. 

\item{{\bf TOPbase, TIPbase, and OPserver}(Feb, 2009)}  \\
{\tt http://cdsweb.u-strasbg.fr/topbase/testop/home.html} provides links to 
databases of atomic data from the Iron Project and Opacity 
Projects, with an emphasis on data for astrophysically abundant ions (Z 
$\leq$ 26).  
                          
\item{{\bf NORAD-Atomic-Data} (Dec, 2013)}   \\
{\tt 
http://www.astronomy.ohio-state.edu/$\sim$nahar/nahar\_radiativeatomicdata/index.html}      
includes more recent data from the Opacity and Iron Projects.    

\item{{\bf  CHIANTI}, version 7.1.4 (May, 2014)}, 
{\tt http://www.chiantidatabase.org/ }        
contains atomic data and programs for computing spectra of astrophysical 
plasmas, with the emphasis on highly-ionized atoms.      

\item{{\bf AtomDB}, version 3.0.0 (Beta 4, Aug, 2014)}, 
{\tt http://www.atomdb.org} 
is an atomic database for X-ray plasma spectral modeling, with the main emphasis 
being the modeling of collisional plasmas.     

\item{{\bf The Vienna Atomic Line Database (VALD3), }(Feb, 2014)}\\
{\tt http://vald.astro.uu.se} is a database that aims to compile complete lists 
of spectral line parameters relevant to the interpretation of stellar 
atmospheric spectra.

\item{{\bf The BIBL database, version 1.58.5} (March, 2014)} \\
{\tt http://das101.isan.troitsk.ru/bibl.htm}
is a comprehensive bibliographic database of experimental and theoretical papers 
on atomic spectroscopy, with an emphasis on papers published since 1983.       
           
\item{{\bf The Virtual Atomic and Molecular Data Centre (VAMDC)}} \\
{\tt http://www.vamdc.eu} provides a uniform access to a large number of atomic 
and molecular databases related to astrophysics.

\end{description}    

\section{Data Needs}                                                                  

Atomic data needs have been summarized in white papers from the AAS Laboratory 
Astrophysics Working Group (now LAD) [\cite{LAW_AAS}]. In this section, a few 
important atomic data needs either for upcoming missions or in support of recent 
analyses of astrophysical spectra are highlighted.
                    
New ground-based and space-based infrared (IR) 
spectrographs such as the Stratospheric Observatory for Infrared Astronomy 
(SOFIA), the Apache Point Observatory Galactic Evolution Experiment (APOGEE), 
and the Cryogenic High-resolution Infrared Echelle Spectrograph (CRIRES) have 
increasing demands for atomic data. The wavelength calibration above 2.5~$\mu$m 
is particularly problematic and new wavelength standards in either atomic or 
molecular species are needed. Measured oscillator strengths in the IR are 
limited to Fe~I, Ti~I, and a handful of lines for other species. Additional data 
are particularly needed in the H-band (around 1.5~$\mu$m). The measurement of 
these oscillator strengths is challenging due to the high excitation of the 
upper energy level of many of the transitions giving lines in this wavelength 
region. 
       
Smith \& Brickhouse [\cite{Smith_AMO}] summarize atomic data 
needs in X-ray astronomy. The interpretation of data from the Chandra X-ray 
Observatory's Low-Energy Transmission Grating (LETG) has been hampered by the 
lack of accurate wavelengths and collisional cross sections in the wavelength 
region from 50~\AA\ to 150~\AA. Spectra from the ASTRO-H SXS microcalorimeter, 
to be launched in 2015, are also likely to be affected. Also required is 
the improvement of the estimation of uncertainties in atomic data.
            
The astrophysical models used to measure the chemical abundances, evolution, and 
physical conditions in many astronomical objects require line identifications, 
wavelengths, oscillator strengths, collision strengths, photoionization 
cross sections, and recombination rate coefficients. Although high accuracy 
experimental wavelengths and oscillator strengths are frequently available for 
lines from low ionization stages, the measurement of oscillator strengths become 
more challenging for higher ionization stages of abundant elements. 
Improved theoretical calculations of moderate accuracy are thus needed, as are 
calculations of photoionization and collisional cross-sections. The large 
differences in abundances obtained from the collisional excitation lines of 
O~III and recombination lines of O~II observed in planetary nebulae has 
partially been resolved by improved calculations [\cite{2012Palay.OIII}] and 
similar calculations for other ions are needed.

One notable example of the influence of new atomic data is the composition and 
opacity of the sun. A new analysis of elemental abundances in the solar spectrum 
resulted in reductions of 30-50~\% in CNO abundances [\cite{Asplund-ARAA}], but 
these revised abundances cannot be reconciled with standard models of the 
stellar interior using helioseismology. A potential resolution to this conflict 
would come from an increase in 15~\% in the opacity of the solar plasmas. A
new laboratory measurement of the iron opacity at conditions similar to those at 
the solar radiation/convection zone boundary provides an indication of this 
[\cite{Bailey-Nature}], with an opacity up to 4 times higher than predicted from 
atomic structure calculations. Since the principal contributors to the opacity 
are photo-excitation and photoionization, the new measurement indicates
that improved calculations are needed for these processes for many abundant 
elements in the sun. Recent calculations for Fe~XVII [\cite{9036TP}], show an 
increase of 20~\% in opacity for this ion and new calculations for other iron 
ions are in progress.           

\section{Notes for References}
The references are identified by a running number. This refers to the general 
reference list at the end of this report, where the literature is ordered
alphabetically according to the first author. Each reference contains one or 
more code letters indicating the method applied by the authors, defined as 
follows:
                                                                         
\begin{description}
\vspace{0.1in}
\item{{\bf THEORETICAL METHODS:}}       
                                                 
\vspace{0.1in}
\begin{tabular}{ll}
{\bf Q: } quantum mechanical calculations. &
{\bf QF: } Calculations of forbidden lines.          \\   
\end{tabular}
\vspace{0.1in}
\item{{\bf EXPERIMENTAL METHODS:}}

\vspace{0.1in}
\begin{tabular}{ll}     
{\bf CL:} New classifications &                            
{\bf EL:}  Energy levels.    \\                   
{\bf WL:}  Wavelengths.       &                                                                 
{\bf HFS:} Hyperfine structure. \\
{\bf IS:}  Isotope structure. &
{\bf L:}   Lifetimes. \\
{\bf TE:} Experimental oscillator strengths. &
{\bf PI:}  Photoionization cross-sections. 
\\
\end{tabular}
\vspace{0.1in}
\item{{\bf OTHER:}} 

\vspace{0.1in}
\begin{tabular}{lll}
{\bf CP: } Data compilations. &
{\bf R: }  Relative values only. &
{\bf F: } Forbidden lines. \\
\end{tabular}

\end{description}              

\vspace{3mm}
\begin{flushright}
Gillian Nave  \\       
{\it chair of Working Group}
\end{flushright}

\section{References on lifetimes and oscillator strengths}\label{TP_refs}

\bibliographystyle{iau}  
\begin{minipage}[t]{1.8in}
Ag I :  \cite{9348TP} \\

Al II :  \cite{9060TP} \\
Al IV :  \cite{9349TP} \\
Al X: \cite{2014Aggarwal.AlX} \\
Al XI :  \cite{9015TP} \\
	Al XII :  \cite{8954TP,9152TP} \\

	Ar II :  \cite{9392TP,9376TP,9448TP} \\
	Ar III :  \cite{9072TP,9003TP} \\
	Ar VI :  \cite{9134TP} \\
	Ar XV :  \cite{9351TP} \\
	Ar XVI :  \cite{2013Nahar.Ar,9245TP} \\
	Ar XVII :  \cite{2013Nahar.Ar,8937TP} \\

	Ba II :  \cite{9210TP} \\

	Be II :  \cite{9210TP,9273TP,9148TP} \\

	C II :  \cite{9221TP} \\
	C III :  \cite{9251TP} \\
	C IV :  \cite{9015TP,9182TP} \\

	Ca I :  \cite{9178TP} \\
	Ca II :  \cite{9464TP,9210TP} \\
	Ca III :  \cite{9172TP} \\
	Ca IX :  \cite{9357TP} \\
	Ca XIV :  \cite{9334TP} \\
	Ca XVIII :  \cite{9161TP,9245TP} \\
	Ca XIX :  \cite{8955TP} \\

	Cd II :  \cite{9210TP} \\

      Ce III : \cite{19069EL,9447TP} \\

	Cl I :  \cite{9165TP} \\
	Cl II-V : \cite{9111TP} \\
	Cl III :  \cite{9117TP} \\
	Cl IV :  \cite{9335TP} \\
	Cl XIV :  \cite{9383TP} \\
	Cl XVI :  \cite{8955TP} \\

	Co XVIII-XXVI : \cite{9314TP} \\
	Co II :  \cite{9217TP} \\

	Cr II : \cite{9013TP,8995TP} \\
	Cr XII : \cite{9252TP} \\
	Cr XXIII : \cite{9064TP} \\
	\end{minipage}
	\begin{minipage}[t]{2in}  
	Cu I : \cite{9348TP,9280TP} \\

	Eu I : \cite{9376TP,9419TP,9046TP} \\
	Eu II : \cite{9225TP,9046TP} \\
	F I : \cite{9339TP} \\

	Fe I :  \cite{9397TP,9034TP,9316TP,9216TP} \\
	Fe II :  \cite{8961TP,9262TP,9034TP,9171TP} \\
	Fe VII :  \cite{9312TP} \\
	Fe VIII :  \cite{8988TP,8965TP}\\
	Fe IX :  \cite{9310TP,8988TP}\\
	Fe X :  \cite{9081TP} \\
	Fe XI :  \cite{9222TP,18347EL,9289TP} \\
	Fe XII :  \cite{9080TP} \\
	Fe XIII :  \cite{9079TP} \\
	Fe XIV :  \cite{2014Aggarwal.FeXIV,9063TP,9213TP} \\
Fe XVI :  \cite{8964TP,9209TP} \\
Fe XVII :  \cite{9215TP,8964TP,9199TP,9036TP} \\
Fe XVIII :  \cite{9089TP} \\
Fe XXII :  \cite{9204TP} \\
Fe XXIII :  \cite{9393TP,9251TP} \\
Fe XXIV  :  \cite{9161TP,9245TP} \\
Fe XXV :  \cite{9175TP} \\

Ga II :  \cite{9268TP} \\
Ga III :  \cite{9132TP} \\

Gd I-II :  \cite{8969TP,9154TP,9259TP} \\

He I :  \cite{9189TP} \\

Hf II :  \cite{9462TP} \\

Hg II :  \cite{9210TP} \\

In I :  O \cite{9143TP} \\

K I :  \cite{9040TP,9128TP} \\
K XVI :  \cite{9383TP} \\
K XVIII :  \cite{8955TP} \\

Kr I :  \cite{9012TP} \\
Kr II :  \cite{9448TP} \\

La I :  \cite{8950TP,9460TP,9309TP} \\
La II :  \cite{9309TP} \\
La III :  \cite{9214TP} \\
	\end{minipage}
	\begin{minipage}[t]{2in}
Li I :  \cite{9015TP} \\

Lu I :  \cite{8957TP} \\
Lu III :  \cite{9170TP} \\

Mg I :  \cite{9178TP} \\
Mg II :  \cite{9210TP} \\
Mg III-XI :  \cite{9078TP} \\
Mg III :  \cite{9411TP,9229TP,9349TP} \\
Mg V :  \cite{9417TP} \\
Mg VIII :  \cite{9174TP} \\
Mg XI :  \cite{8954TP} \\

Mn I :  \cite{18980EL,9135TP,9466TP,9345TP,9463TP} \\
Mn II :   \cite{19113EL,9135TP,9140TP} \\
Mn XII :  \cite{9320TP} \\
Mn X-XIII :  \cite{9111TP} \\
Mn XXIV :  \cite{9064TP} \\

Mo I :  \cite{9151TP} \\
Mo II :  \cite{9087TP} \\
Mo III :  \cite{9389TP} \\

N I :  \cite{9315TP} \\
N II :  \cite{9420TP,9258TP} \\
N V :  \cite{9182TP} \\

Na I :  \cite{9184TP} \\
Na VII :  \cite{19088EL} \\
Na II-X :  \cite{9006TP} \\

Nb II :  \cite{9425TP} \\

Nd I :  \cite{8948TP} \\

Ne I :  \cite{9250TP,9126TP,9398TP,8992TP} \\
Ne II :  \cite{9448TP} \\
Ne IV :  \cite{9322TP,2014Nahar.NeIV} \\
Ne V: \cite{2013Dance.NeV} \\
Ne VI :  \cite{9174TP} \\
Ne VII :  \cite{9091TP,9440TP,9251TP} \\

Ni I :  \cite{9463TP,9281TP} \\
Ni II :  \cite{9023TP} \\
Ni X :  \cite{8945TP} \\
Ni XI :  \cite{9319TP} \\
Ni XIV :  \cite{9096TP} \\
Ni XV :  \cite{9340TP,9232TP,9077TP,9098TP} \\
	\end{minipage}
\newpage
	\begin{minipage}[t]{2in}
Ni XVII :  \cite{8987TP} \\
Ni XXI :  \cite{9375TP} \\
Ni XXVI :  \cite{9007TP} \\
Ni XXVII :  \cite{9179TP} \\

O I :  \cite{9307TP} \\
O II :  \cite{9049TP} \\
O III :  \cite{9307TP,2012Palay.OIII,2014Storey.OIII} \\
O IV :  \cite{9350TP,9174TP} \\
O V-VI :  \cite{9255TP} \\
O VI :  \cite{8963TP,9182TP} \\

P VIII-X :  \cite{9002TP} \\
P XIV :  \cite{8954TP} \\

Pb III :  \cite{8962TP} \\

Re I :  \cite{9018TP} \\
Re II :  \cite{9228TP} \\

Rh I :  \cite{9449TP} \\
Rh II :  \cite{9224TP,8989TP} \\

Ru I :  \cite{8939TP} \\

S II : \cite{2014Kisielius.SII} \\
S VII :  \cite{9065TP} \\
S VIII : \cite{9039TP} \\
S IX,XI :  \cite{9002TP} \\
S IX-XII : \cite{8934TP} \\
S XIII : \cite{9198TP,9020TP} \\
S XV :  \cite{8954TP} \\
	\end{minipage}
	\begin{minipage}[t]{2in}
Sc I :  \cite{9374TP} \\
Sc II :  \cite{9285TP} \\
Sc III :  \cite{8947TP,9019TP} \\
Sc III-XXI :  \cite{9017TP} \\
Sc XX :  \cite{8955TP} \\

Si I :   \cite{8981TP} \\
Si II: \cite{2014Aggarwal.SiII}\\
Si III :  \cite{9415TP} \\
Si IV :  \cite{19014EL} \\
Si IX :  \cite{9321TP} \\
Si V-XIII :  \cite{8951TP} \\
Si VII :  \cite{9305TP} \\
Si VIII :  \cite{9011TP,9002TP} \\
Si X :  \cite{9174TP} \\
Si XII :  \cite{9245TP} \\

Sm I :  \cite{9193TP,9218TP} \\
Sm II :  \cite{9047TP} \\

Sn I :  \cite{18891EL,9050TP} \\
Sn III :  \cite{9068TP} \\

Sr II :  \cite{9210TP,8986TP} \\

Tb I :  \cite{9226TP} \\

Te II-III :  \cite{9160TP} \\

Th II :  \cite{8978TP} \\
	\end{minipage}
	\begin{minipage}[t]{2in}
Ti I :  \cite{9158TP,2015Nahar.TiI} \\
Ti II :  \cite{9372TP,9363TP} \\
Ti IV: \cite{19014EL} \\
Ti VI :  \cite{9205TP} \\
Ti VII :  \cite{9104TP} \\
Ti VII-X :  \cite{9111TP} \\
Ti X :  \cite{9206TP} \\
Ti XIX :  \cite{9130TP} \\
Ti XX  :  \cite{9245TP} \\
Ti XXI :  \cite{9064TP} \\

Tl I :  \cite{9157TP} \\

U I :  \cite{8977TP,9331TP} \\

V I :  \cite{9282TP} \\
V II :  \cite{9217TP} \\
V XII :  \cite{9207TP} \\
V XXII :  \cite{9064TP} \\

W II :  \cite{8949TP} \\

Xe II : \cite{9448TP}\\
Xe VI : \cite{9406TP,9456TP}\\

Y I :  \cite{9421TP} \\
Y III :  \cite{9149TP} \\

Yb I :  \cite{8994TP} \\

Zn I :  \cite{9268TP,9261TP,8960TP} \\
Zn II :  \cite{9210TP,19014EL} \\

Zr II :  \cite{19189EL,9475TP} \\
	\end{minipage}

	\begin{minipage}[t]{3in}
He-like ions: \cite{8952TP,9436TP,9176TP} \\

Li-like ions: \cite{9150TP,9323TP,8944TP,19127EL} \\

Be-like ions: \cite{2015Aggarwal.Belike,9201TP,9127TP,9136TP,19065EL} \\

B-like ions: \cite{9256TP,9139TP,9062TP,9237TP,9145TP} \\

C-like ions: \cite{9269TP,8993TP,9212TP,9330TP,9241TP,9396TP,9391TP} \\

N-like ions: \cite{9291TP,9242TP} \\
	\end{minipage}
	\begin{minipage}[t]{3in}
O-like ions: \cite{9186TP} \\

Mg-like ions: \cite{9373TP,9326TP} \\

Ne-like ions: \cite{9234TP,9208TP,9386TP} \\

Si-like ions: \cite{9094TP} \\

F-like ions: \cite{9164TP,9279TP} \\

Neutral Li-Ar :  \cite{9101TP} \\
\end{minipage}

\end{document}